# Applications of Controlled-Flow Laser-Polarized Xenon Gas to Porous and Granular Media Study

R. W. Mair[1,2], R. Wang[1,2], M. S. Rosen[1], D. Candela[3]

D. G. Cory[2], and R. L. Walsworth[1]

[1] Harvard-Smithsonian Center for Astrophysics, Cambridge, MA 02138, USA.

[2] Massachusetts Institute of Technology, Cambridge, MA, 02139, USA.

[3] University of Massachusetts, Amherst, MA, 01003, USA

**Corresponding Author:**

Ross Mair

Harvard Smithsonian Center for Astrophysics,

60 Garden St, MS 59,

Cambridge, MA, 02138,

USA

Phone: 1-617-495 7218

Fax: 1-617-496 7690

Email: rmair@cfa.harvard.edu

# ABSTRACT

We report initial NMR studies of continuous flow laser-polarized xenon gas, both in unrestricted tubing, and in a model porous media. The study uses Pulsed Gradient Spin Echo-based techniques in the gas-phase, with the aim of obtaining more sophisticated information than just translational self-diffusion coefficients. Pulsed Gradient Echo studies of continuous flow laser-polarized xenon gas in unrestricted tubing indicate clear diffraction minima resulting from a wide distribution of velocities in the flow field. The maximum velocity experienced in the flow can be calculated from this minimum, and is seen to agree with the information from the complete velocity spectrum, or motion propagator, as well as previously published images. The susceptibility of gas flows to parameters such as gas mixture content, and hence viscosity, are observed in experiments aimed at identifying clear structural features from echo attenuation plots of gas flow in porous media. Gas-phase NMR scattering, or position correlation flow-diffraction, previously clearly seen in the echo attenuation data from laser-polarized xenon flowing through a 2 mm glass bead pack is not so clear in experiments using a different gas mixture. A propagator analysis shows most gas in the sample remains close to static, while a small portion moves through a presumably near-unimpeded path at high velocities.

# INTRODUCTION

NMR techniques are commonly used as non-invasive methods for the study of porous materials. By detecting the [1]H signal from water-saturated rocks and model systems, structural information including the surface-area-to-volume ratio ($S/V_p$) [1-3] and average pore or compartment size [4-6], as well as visualizations of fluid transport under flow [7,8] can be obtained. Usually, these methods study the motion of water spins, and the Pulsed Gradient Spin Echo (PGSE) technique has become a powerful tool for this purpose [9,10]. Except in cases of very small pores or very high fluid flow rates, however, spin relaxation quenches the NMR signal before water molecules can diffuse across even one pore.

We have recently extended these familiar diffusion NMR techniques to porous media imbibed with a gas rather than a liquid, yielding the time-dependent gas diffusion coefficient *D(t)* [11-14]. The spin $\frac{1}{2}$ noble gases ([3]He and [129]Xe) are ideal for such studies, given their high diffusion coefficients, inert nature, low surface interactions, and the ability to tailor the diffusion coefficient to some extent by controlling the gas pressure in the sample. Using [129]Xe gas as the observation spin, with a diffusion coefficient ~ 3 orders of magnitude higher than that of water ($5.7 \times 10^{-6}$ m$^2$ s$^{-1}$ at 1 bar pressure [11]), it has been possible to extend the porous length scales over which diffusion is observed by over an order of magnitude, probing pore length scales on the order of millimeters [12-14]. The advent of the spin-exchange optical pumping technique now allows the production of large, non-equilibrium spin polarizations (~10%) in samples of noble gases, allowing gas phase samples to be created with equivalent magnetization density to that of water when place in fields ~ 1 T [15]. As a result, the technique has become popular in medical imaging, especially as a non-invasive and high-resolution probe of the human and animal lung space [16,17]. We believe, however, that laser-polarized gas NMR also has a vital role to play in materials science investigations.

In addition to the numerous potential applications in porous media, one particular area of interest involves the study of granular systems. NMR is fast becoming a popular tool for studying granular systems due to its ability to non-invasively probe the three-dimensional structure of the opaque system during motion. However, all current experiments have focussed on observing [1]H NMR signals from the granular particles themselves, rather than the surrounding gas [18-23]. In order to facilitate this and other areas of gas-phase NMR study, we have built and begun testing a [129]Xe polarization system to provide a continuous supply of laser-polarized xenon, currently deliverable in a constant, controlled flow mode. In its final configuration, the system will also provide gas in a repeated, batch-delivery mode, supplying multiple shots of polarized xenon of a reproducible pressure and volume. In this paper, we report initial results from this gas delivery system in the constant flow mode.

## MATERIALS AND METHODS

It is well known that the NMR echo signal observed in a PGSE experiment has a Fourier relationship to the probability of spin motion – the so-called displacement propagator, which can be thought of as a spectrum of motion. The echo signal, $E$, obtained in a PGSE experiment can thus be written as [10]:

$$E(\mathbf{q},t) = \int \overline{P}_s(\mathbf{R},t) \times \exp[i2\pi\mathbf{q} \cdot (\mathbf{R})]d\mathbf{R} \qquad (1)$$

where $\overline{P}_s(\mathbf{R}, t)$ is the ensemble average displacement propagator, or the probability of a spin having a displacement $\mathbf{R} = \mathbf{r'} - \mathbf{r}$ proceeding from any initial position $\mathbf{r}$ to a final position $\mathbf{r'}$ during the 'diffusion time' $t$ (often referred to as $\Delta$ in the literature). $\mathbf{q}$ is the wavevector of the magnetization modulation induced in the spins by a field gradient pulse of strength $g$ and pulse duration $\delta$. The magnitude of $\mathbf{q}$ is $\gamma\delta g/2\pi$, where $\gamma$ is the spin gyromagnetic ratio. Consequently, the Fourier transform of $E$ with respect to $\mathbf{q}$ yields an image of $\overline{P}_s$. In the limit of small $\mathbf{q}$, $\overline{P}_s$ is a Gaussian, and it can be shown that the spins undergoing motion will produce an echo with a phase factor $\exp[i2\pi q v t]$ where $v$ is the velocity of the spins, while the echo is attenuated by a factor $\exp(4\pi^2 q^2 D(t) t)$ [10], where $D(t) = \langle [\mathbf{r'}-\mathbf{r}]^2 \rangle / 6t$ is the time-dependent diffusion coefficient describing incoherent random motion of spins in the pore space.

Outside the small $q$ limit, $\overline{P}_s$ is no longer Gaussian. As such, the signal attenuation curve ($q$-plot), is no longer linear with $g^2$, even for single-component diffusion. When $t$ is large enough for spins to completely sample the restrictions of a given pore, $\overline{P}_s$ reduces to the pore spin density function, $\rho(\mathbf{r'})$, and the echo attenuation becomes the Fourier power spectrum of $\rho(\mathbf{r'})$ [5,24]. Specifically, for spins restricted in an open-pore system such as bead packs, the $q$-plot will exhibit a sinc modulation, with a minimum before rising to a maximum at a value of $q$ that corresponds to the reciprocal of the bead diameter. This phenomenon has come to be termed "NMR diffusive diffraction" or "NMR scattering". Other effects are manifested in one-dimensional restricted systems, and when flow is present [24].

Velocity measurements may be obtained from PGSE-style experiments directly from the displacement propagator. The complex signal from each spectrum is Fourier transformed with respect to $q$ to yield the average propagator for those spins. The phase factor accumulated from coherent flow during $t$ manifests itself as an offset of the propagator from zero, which yields the average velocity experienced by the spins in the sample during $t$. Using the Fourier encoding velocity method, after transformation, the velocity is calculated from [10]:

$$v = (2\pi g_{steps} k_v) / (q_{fft} \gamma \delta t g) \qquad (2)$$

where $g_{steps}$ = the number of gradient values used (including zero), $k_v$ = the offset in digital points from zero of the maximum of the velocity spectrum, and $q_{fft}$ is the number of points Fourier transformed (including those used for zero-filling) with respect to $q$.

Laser polarization was achieved by spin exchange collisions between the $^{129}$Xe atoms and optically pumped rubidium vapor. The gas mixture used contained either 92% xenon and 8% nitrogen, or 5% xenon, 10% nitrogen and 85% helium. The xenon was optically pumped at ~ 110° C for approximately 15 minutes, using circularly polarized light at 795 nm from a 15 W fiber-coupled diode laser array [Optopower Corp, Tucson, AZ]. After optical pumping, the polarization chamber was opened to a previously evacuated sample in the NMR magnet, connected by thin teflon tubing. The end of the sample was connected to a Mass Flow Controller [MKS Instruments, Methuen, MA] which had a capacity for regulating flows of 0 to 1000 cm$^3$/min. A vacuum pump completed the circuit. Under the influence of the pump and the Mass Flow Controller, xenon flowed continuously from the supply bottle through the polarization chamber, then to the sample in the NMR magnet, and finally to the pump. With a suitable supply of xenon gas mixture, stable flows could be maintained for many hours. The samples used for the experiments reported here include 0.125 inch inner diameter flexible teflon tubing which was looped through the RF coil 6 times, and a glass cell that contained 2 mm glass beads. Slower gas flow rates resulted in higher initial xenon polarization as the spins spent longer in the polarization chamber, but suffered greater $T_1$ relaxation in the teflon tubing before reaching the RF coil. Maximum signal was obtained for flow rates ~ 300 - 400 cm$^3$/min. This higher signal at higher flow rates also permitted fewer signal averaging scans.

For experiments on laser-polarized gas flow in tubes without obstructions or glass beads, background gradient compensation was not necessary, and $T_2$ was long. Hence, the simple PGSE-related methods were used, although modifications for use with laser-polarized gases were incorporated - specifically, the use of low flip angle excitation pulses and the removal of 180° pulses, i.e., the Pulsed Gradient Echo method[11]. For gas flow in porous samples, we used a modified stimulated echo sequence incorporating alternating bi-polar diffusion encoding gradient pulses (PGSTE-bp) which served to cancel out the effect of the background gradients while applying the diffusion encoding gradient pulses [14, 25,26]. The sequence is illustrated in Fig. 1, where the labeled timing parameters correspond to the description in the previous section. All experiments, were carried out using a Bruker AMX2 - based spectrometer (Bruker Instruments Inc., Billerica, MA) interfaced to a 4.7 T magnet. This system is equipped with a 12 cm ID gradient insert (Bruker) capable of delivering gradient pulses of up to 26 G/cm. We employed an Alderman-Grant-style RF coil, tuned to 55.35 MHz for $^{129}$Xe observation [Nova Medical Inc., Wakefield, MA]. Acquisition parameters are noted in the figure captions.

## RESULTS AND DISCUSSION

Initial tests of the continuous flow laser-polarized xenon facility involved measurements of the gas velocity at different flow rates while experiencing unobstructed flow in straight tubes or pipes. This allowed testing of the apparatus, as well as ensuring the effectiveness of velocity mapping on flowing gas. In order to increase the gas velocity while maintaining tolerable mass flow rates, we conducted experiments of gas flow in narrow teflon tubing with a 0.125 inch (~ 3 mm) internal diameter. The tubing was looped back and forth through the RF coil 6 times, giving flow in opposite directions in each of three tubes. We have previously reported two-dimensional velocity and diffusion images from this experimental setup [27]. Imaging was accomplished by combining a PGE sequence for flow encoding and gradient-recalled echo imaging sequence. Average velocities of ~ ± 60 and ± 120 mm/s were observed in most pixels at mass flow rates of 100 and 200 cm$^3$/min, although some velocity variation within each tube was apparent, especially at the higher flow rate.

Using the same experimental apparatus, simpler, non-localized measures of the flow field were made using the Pulsed Gradient Echo method [11]. The net mass flow of xenon through the RF coil was zero, as the tubing looped back and forth three times in each direction and exited the coil on the same side as it entered. However, repeated RF pulsing of the laser-polarized gas results in a loss of magnetization that is not recovered. Therefore, a net flow of magnetization occurs in one direction, resulting in an observable averaged flow in this experimental apparatus. Signal attenuation plots as a function of $q$, obtained at four different flow rates are shown in Fig. 2a). The monotonic decrease in the $q$-plot at the lowest flow rate gives way to a sinc attenuation profile resulting from the distribution of velocities experienced in the sample. This phenomenon, which does not infer structural information, but rather, details of the flow field, has recently been termed "displacement diffraction" [24]. This effect was first observed in liquid laminar flow 30 years ago by Hayward et al [28], and has also been seen in samples suffering convection [29]. The first minimum occurs at the reciprocal of the maximum spin displacement during the flow encode time, allowing an estimation of the maximum velocity being experienced in the flow field.

Fourier transforming these $q$-plots gives the velocity distribution spectra, or motion propagators for the fluid flow - these are shown in Fig. 2b). The uniform flow indicated by a Gaussian propagator at low flow rates gives way to something closer to a hat function at the 200 cm$^3$/min flow rate, with maximum velocities of ~ ± 175 - 200 mm/s, as was observed in some pixels in the velocity images from this apparatus, reported previously [27]. However, throughout the sample, there is roughly equal probability of all velocities between these two limits being observed. It should be noted the propagators for 100 and 150 cm$^3$/min flow rates are not as well defined as those at higher flow rates due to the acquisition of only 16 $q$ points at these flow rates, rather than the 64 used at the higher flow rates. This will account for some of the variation in appearance. Conversely, the higher flow rates could have also reduced the amount of magnetization loss in the sample, and thus the observed flow became more equally weighted to spins moving in both directions. It is constructive to compare the maximum velocities that can be estimated from the minima of the echo attenuation plots in Fig. 2a) with the appearance of the complete velocity spectra in Fig 2b). For the flow rates of 150, 200 and 300 cm$^3$/min; minima are observed at $q$ = 1770, 1330 and 900 m$^{-1}$ respectively. This corresponds to net displacements of 0.56, 0.75 and 1.11 mm, or maximum velocities of ~ 190, 250 and 370 mm/s respectively - which agrees well with that observed from the propagators.

A second test of the continuous flow laser-polarized xenon apparatus involved flowing the gas through a 2 mm glass bead pack, and studying the echo attenuation in a PGSTE-bp spectroscopy experiment. Initial experiments were performed using the 5% xenon gas mixture - the higher net xenon diffusion resulted in a maximum flow encoding time of 0.5 s being used. For a system of uniform packed beads with diameter $b$, diffusion or flow over time scales long enough for spins to traverse multiple pores will result in a "NMR scattering" or "diffusive/flow diffraction" pattern that results from the pore structure, rather than the flow field itself [24]. It is known the echo attenuation maximum that follows the first minimum occurs at $q = b^{-1}$ [24] and it can thus be shown that the first minimum occurs at $q \approx 0.7b^{-1}$.

At the highest flow rate of 1000 cm$^3$/min, a clear diffraction minimum was observed which correlated well with the bead size ($q = 323$ m$^{-1}$ $\Rightarrow b = 2.17$ mm). This data was reported in ref [27].

For the current experiments, a 92% xenon gas mixture was used, thus allowing the use of longer flow encode times as a result of the lower net xenon diffusion coefficient in this mixture. The data we report used $t = 1.5$ s; the echo attenuation plots are given in Fig. 3a), along with the data obtained at 1000 cm$^3$/min and $t = 0.5$ s reported previously [27]. The initial monotonic decrease at the lowest flow rate tends towards the sinc attenuation profile expected at the higher flow rates, however the trending is less uniform, and a single clear diffraction minimum is not apparent in this data. A shallow minimum, albeit at a considerably higher value of $q$, is observed at 300 cm$^3$/min, while a very sharp, narrow minimum is seen at 400 cm$^3$/min, very close to the expected value for this bead pack. However, at 600 cm$^3$/min, the minimum has disappeared, and the signal attenuation plot is featureless

As a clue to the cause of this phenomenon, velocity spectra are provided in two cases in Fig. 3b). The spectrum for the 5% xenon mixture at 1000 cm$^3$/min shows a single, very broad peak, with its maximum at ~ 5 mm/s. In addition, because of its asymmetric shape, a large fraction of spins can be seen to be moving at velocities of 8 - 12 mm/s, indicating that many spins will traverse more than one pore during the 0.5 second observation time. This behavior results in the clear definition of the diffraction minima seen in ref. [27]. However, for the 92% xenon mixture at 400 cm$^3$/min, a much narrower velocity peak is seen. It is centered at a velocity of ~ 3 mm/s, with very few spins experiencing velocities above 5 mm/s. This indicates that significant gas hold-up is occurring within the bead pack and the majority of the spins are moving much slower than the expected average velocity based on the mass flow rate, implying many spins are not traversing multiple pores during the measurement. However, a low but extremely broad peak of velocities is seen from 20 - 40 mm/s, most likely indicating that while most of the gas remains close to static in the sample, a small jet of high speed gas is rapidly moving through a portion of the sample - most likely the middle. While the reduced xenon diffusion in the 92% xenon mixture will result in reduced dispersion, and hence a narrowing of the velocity distribution in this mixture, we believe the appearance of the broad high-velocity peak must be related to changes in the viscosity or other parameters dependent on the gas content, as no other variables in the sample or NMR experiment were modified.

# CONCLUSIONS

Gas-phase NMR diffusion and flow measurements are a powerful tool for the study of porous media and gas-flow dynamics, as well as providing the potential for the first time to study the gas motion in granular systems. In the current work, we have carried out initial tests of a new continuous flow laser-polarized xenon gas delivery system, studying unrestricted xenon flow through straight tubing, as well as identifying structural parameters and the effect of gas mixture on flow through a 2 mm bead pack.

These initial experiments with the continuous flow laser-polarized xenon delivery apparatus show we have the ability to study (and image) high velocity gas flows (20 - 200 mm/s), permitting the study of fluid flow rates much higher than can generally be obtained with liquids. For xenon flow in multiple loops of narrow 0.125 inch diameter tubing, we observed clear "displacement diffraction" minima, which occurred at lower values of $q$ as the flow rate was increased. This form of NMR diffraction occurs due a wide distribution of velocities in the flow field, not any structural feature. This distribution of velocities will also result in Taylor dispersion effects in this sample, an effect that was previously indicated by higher apparent xenon diffusion coefficients measured in the imaging experiments [27]. In addition, the structure of the signal attenuation plot permits an estimation of the maximum velocity in the system, without resorting to the full propagator of motion.

We had previously presented the first gas-phase NMR scattering, or position correlation flow-diffraction data, exhibiting flow-enhanced structural features from laser-polarized xenon flowing through a 2 mm glass bead pack [27]. Additional experiments using a more concentrated xenon gas mixture, resulted in a loss of structural features in the $q$-plot at higher flow rates, as most gas remained close to static while a small region of gas moved through the sample at high velocity in an almost unimpeded pathway.

Flowing laser-polarized noble gas NMR has the potential to be a significant tool in porous and granular media study, as well as in gas fluid dynamics. The much higher gas flow rates permissible allow NMR scattering techniques to probe much longer regular length scales than could be probed with liquids. The ability to observe a NMR sensitive gas with high SNR will also allow future studies of the gas dynamics in granular media, especially gas-fluidized systems. While powerful, we note that gas velocity imaging has a diffusive-attenuation-induced velocity resolution limit of about 1 mm/s, in a similar manner to the diffusive-limited gas imaging resolution limit in MRI applications [27].

## ACKNOWLEDGEMENTS

RWM gratefully acknowledges scientific discussions with Profs. Paul Callaghan and Ken Packer at the 6th International Conference on Magnetic Resonance in Porous Media, who helped clarify the phenomena observed in Fig. 2a). This work was supported by NSF grant CTS-9980194, NASA grant NAG9-1166 and the Smithsonian Institution.

# FIGURE CAPTIONS

Fig. 1. Pulse sequence diagram for the Pulsed Gradient Stimulated Echo with alternating bi-polar gradient pulses (PGSTE-bp), as used in this work. The diffusion encoding gradient pulses, of length $\delta$ and strength $g$, are shown in gray, while crusher gradients are shown in black. The diffusion time is denoted $t$, and the total diffusion encoding time is $T$. See text for further description.

Fig. 2. (a) Echo attenuation plots, $\ln(S(q)/S(0))$ vs $q$ ($q = \gamma \delta g / 2\pi$), from PGE spectroscopy experiments for laser-polarized xenon flowing through 6 back-and-forth loops of 0.125 inch diameter teflon tubing. This averaged data includes contributions from spins moving in both directions in the RF coil. (b) Velocity distribution spectra resulting from the Fourier transform of the data in (a). Acquisition parameters: ~ 33° RF excitation pulse was used to conserve magnetization; $\delta = 1$ ms, $t = 3$ ms, $g$ was stepped linearly from 0 to 20.5 G/cm in 16 steps or 24.6 G/cm in 64 steps. The repetition rate was varied with the flow rate from 2 to 0.5 s as the flow rate increased.

Fig. 3. (a) Echo attenuation plots, $\ln(S(q)/S(0))$ vs $q$ ($q = \gamma \delta g / 2\pi$), from PGSTE-bp spectroscopy experiments for laser-polarized xenon flowing through a pack of 2 mm glass beads. The 92% xenon gas mixture was used for most experiments, allowing $t = 1.5$ s without complete diffusive signal attenuation. The data series with $t = 0.5$ s was acquired using the 5% xenon mixture, and is reproduced from ref. [27]. (b) Velocity spectra resulting from the Fourier transform of the 400 and 1000 cm$^3$/min data in (a). Acquisition parameters: 90° RF excitation pulse, $\delta = 750$ µs, $g$ was incremented linearly in 32 steps from 0 to 3.78 G/cm. The repetition rate was varied with the flow rate to ensure enough time for freshly polarized xenon to flow completely into the sample chamber, ranging from 5 to 20 seconds.

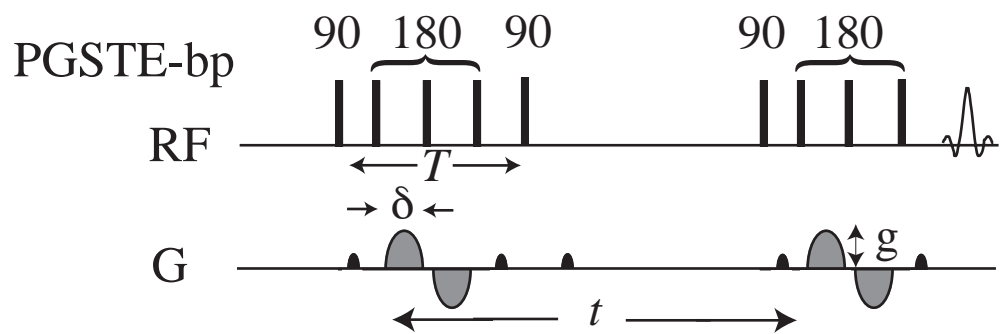

**Figure 1**

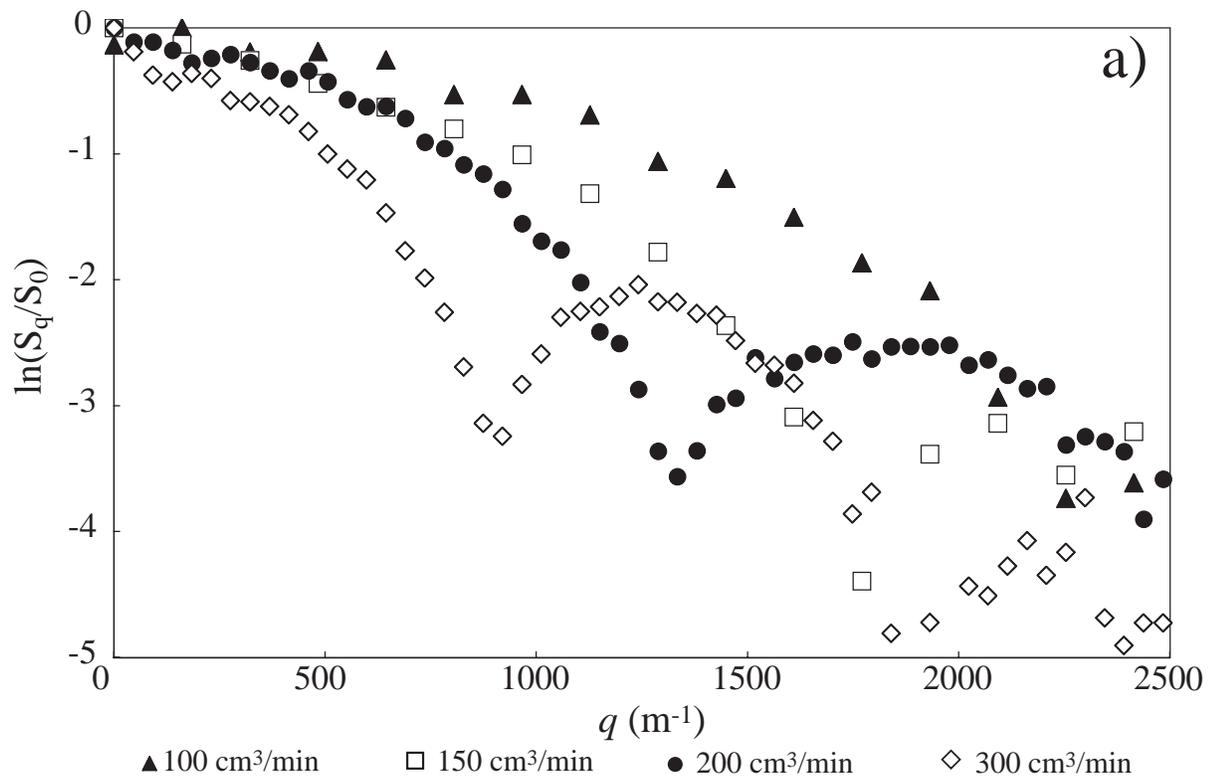
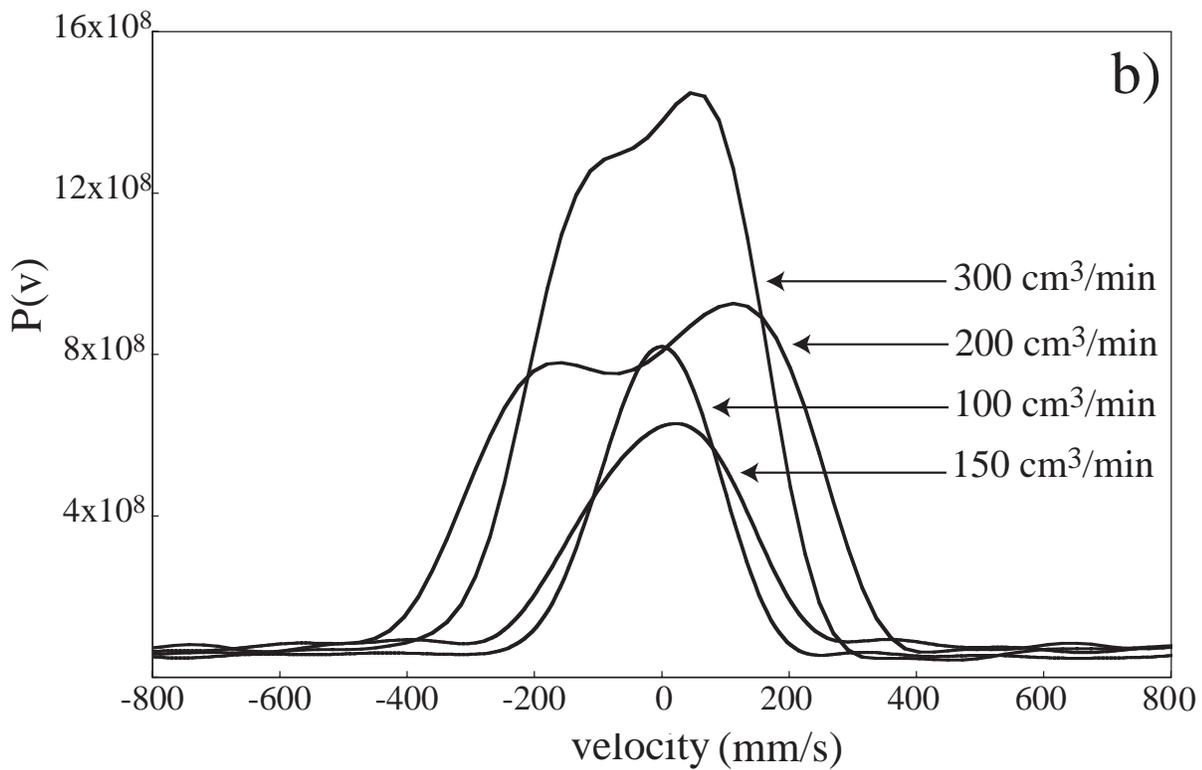

**Figure 2**

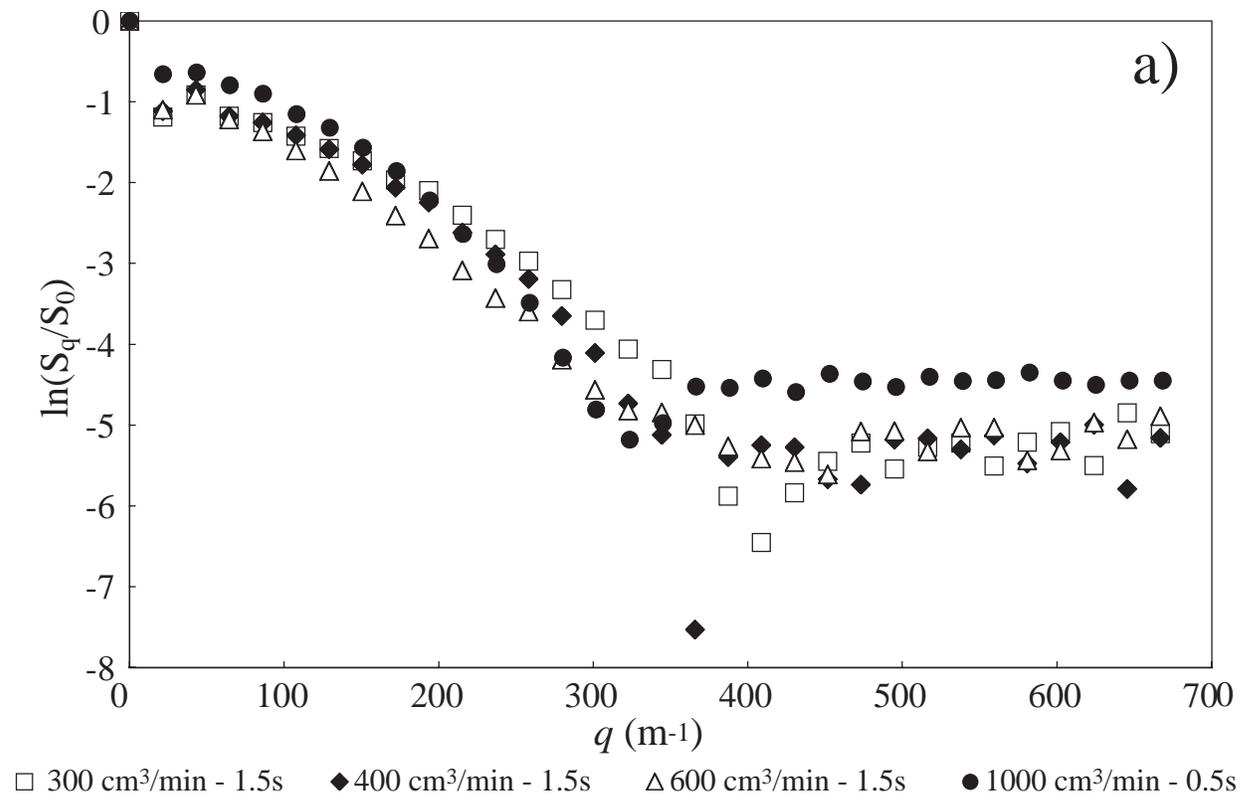
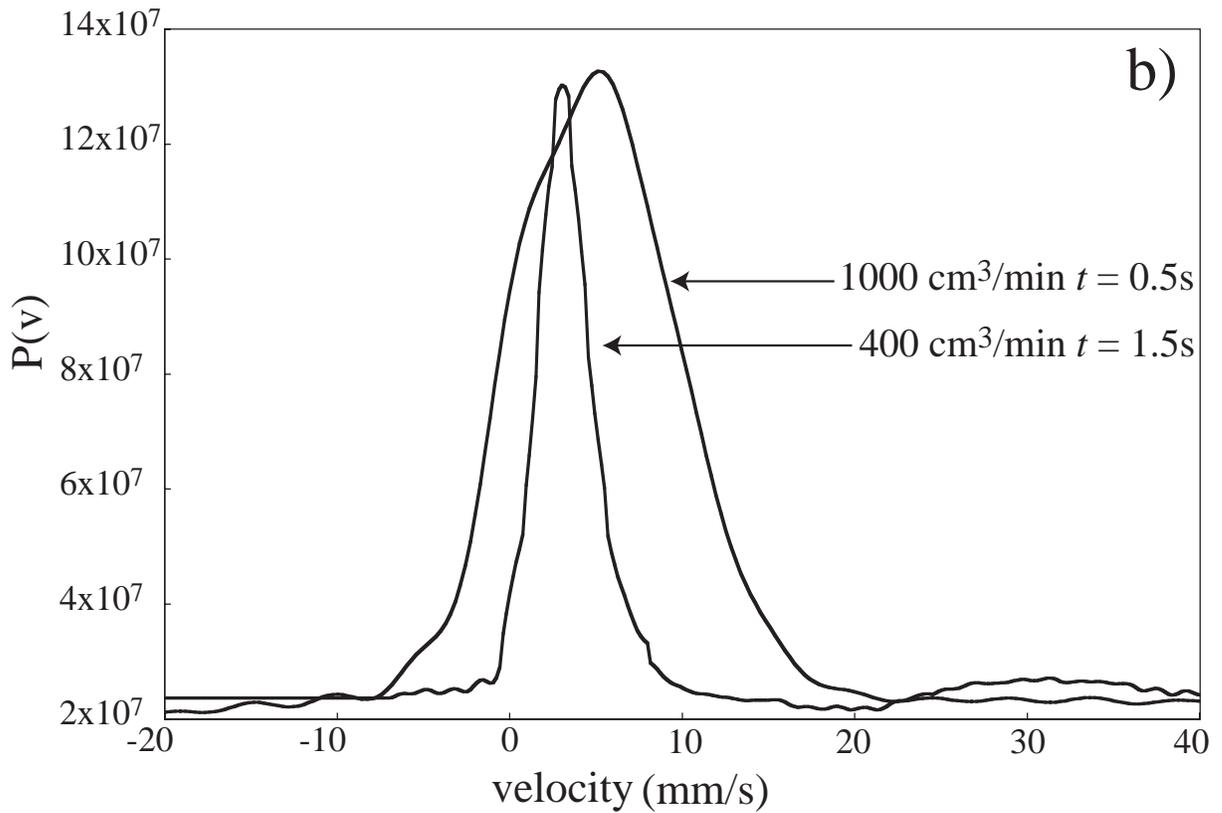

**Figure 3**